\documentclass[11pt,twoside]{article}

\usepackage{asp2006}
\usepackage{natbib}
\usepackage{graphicx}

\begin{document}
\title{Pseudo-global Fitting of Gapped Helioseismic Data}   
\author{Stephen Fletcher and Roger New}   
\affil{Sheffield Hallam University, Sheffield, UK}    
\author{William Chaplin and Yvonne Elsworth}   
\affil{University of Birmingham, Birmingham, UK}    

\begin{abstract} 
Mode fitting or ``peak-bagging" is an important procedure in
helioseismology allowing one to determine the various mode
parameters of solar oscillations. We have recently developed a new
``pseudo-global" fitting algorithm as a way of reducing the
systematic bias in the fits of certain mode parameters that are seen
when using ``local" fitting techniques to analyse ``sun-as-a-star"
p-mode data. This new algorithm has been designed specifically to
gain the advantages of fitting the entire power spectrum, while
retaining the efficiency of local fitting techniques.

Using simulated data with a full fill we have previously shown that
the pseudo-global routine reduces the bias in estimates of the
frequencies and asymmetries and in the estimates of the solar
background when compared with a traditional fitting technique. Here
we present results that show that the pseudo-global routine is also
effective in reducing bias in the parameter estimates when the
time-series has significant gaps. As such we are now able to employ
the routine in order to fit ground based helioseismic data such as
that collected by the Birmingham Solar Oscillations Network (BiSON).
\end{abstract}



\section{Introduction}

Being able to determine accurate estimates of the various solar mode
parameters is an important goal in helioseismology. Using inversion
techniques frequencies of the oscillation modes can be used to
constrain estimates of the sound speed and density in the solar
interior, while the rotational splitting of the modes can determine
the internal rotation rate.

Over the years the quality and quantity of helioseismic data has
improved significantly enabling the parameter values to be
constrained with increasingly greater precision. However, with
greater precision comes the need for greater accuracy and this can
only be archived if the models used to fit the data are also
accurate.

For low-degree (low-$\ell$) Sun-as-a-star observations, the most
common method of fitting is to split the power spectrum into a
series of fitting regions centered on $\ell$ = 0/2 and 1/3 pairs.
The modes are then fitted, pair by pair, to determine the parameter
values without the need to fit the entire spectrum simultaneously.

The main advantage of this ``pair-by-pair" fitting method (hereafter
abbreviated PPM) is its computational efficiency, since the number
of parameters being varied is small. However, the cost of this
efficiency is that the model used to fit the data encompasses only
those modes within the fitting region. Hence, any power from modes
whose central frequencies lie outside this region will not be
accounted for. This imperfect match between the fitting model and
the underlying profile of the data can lead to significant biases in
some of the fitted parameters.

In \cite{Fletcher2008a,Fletcher2008b} (hereafter paper's I and II)
we introduced a modified fitting routine that takes into account the
effect of modes that lie outside the fitting regions. This was done
by employing a model which is valid for the entire spectrum.
However, in order to retain computational efficiency only the
parameters associated with those modes within the fitting region
were allowed to vary. Hence we refer to this technique as a
pseudo-global fitting method (hereafter abbreviated PGM).

In this paper we modify the PGM method further in order to fit
spectra made from time series that have significant gaps. An
overview of the changes made are given in
section~\ref{SecTechniques} Using simulated data we show that the
PGM remains a robust fitting strategy. Being able to fit
gap-affected data allows us to test the PGM on ground based data
such as that collected by the Birmingham Solar Oscillations Network
(BiSON). Results of the fits to both the simulated and BiSON data,
along with comparisons with fits from the traditional PPM are given
in Section~\ref{SecResults}

\section{Fitting Techniques}\label{SecTechniques}

The PPM and PGM routines were described at length in Paper's I and
II so only the details of modifications made to treat gap-affected
data are given here. The best way of accounting for gaps is to
convolve the power spectrum fitting model with the spectral window
and this technique was incorporated in both the PPM and PGM
routines.

For the PPM, this is a fairly straight-forward modification.
However, in the case of the PGM the convolution with the spectral
window is a little more complicated. The background is accounted for
by fitting a model to the low and high frequency ends of the power
spectrum, well away from the regimes of the p-modes. Two background
sources need to be considered for the model. The first of these is a
granulation-like component that mimics the solar velocity continuum.
The power spectral density of which may be described by a power-law
model \citep[see][]{Harvey1985}. Secondly, a frequency independent
term, $\sigma_s$, is added to account for the uncertainty in the
velocity measurements caused by photon shot noise. Finally, when
gapped data is being considered, the full model used to fit the
background is convolved with the spectral window, $W$. Hence, the
full power spectral density of the background, $n(\nu)$, is given
by:
\begin{equation}
n(\nu) = \left[\frac{2\sigma_g^2\tau_g}{1+(2\pi\nu\tau_g)^2}+
\sigma_s\right] \bigotimes W \label{Harvey}
\end{equation}
where the values of $\tau_g$ and $\sigma_g$ give the lifetime and
amplitude of the granulation profile. Depending on how low in
frequency one fits, further terms may be required to account for
meso- and super-granulation.

Once the background fit is performed the main mode-fitting process
can begin. A starting model is generated for the entire spectrum and
is derived from the results of an initial run of the pair-by-pair
method and the results of the background fit. For gapped data this
model is then convolved with the spectral window and fitted to the
data, allowing only the parameters associated with the modes within
the fitting window to vary.

Fig. 1 gives an updated flow diagram of the main steps involved in
performing the PGM when fitting gap affected data.

\begin{figure}
\centerline{\includegraphics[width=4.0in]{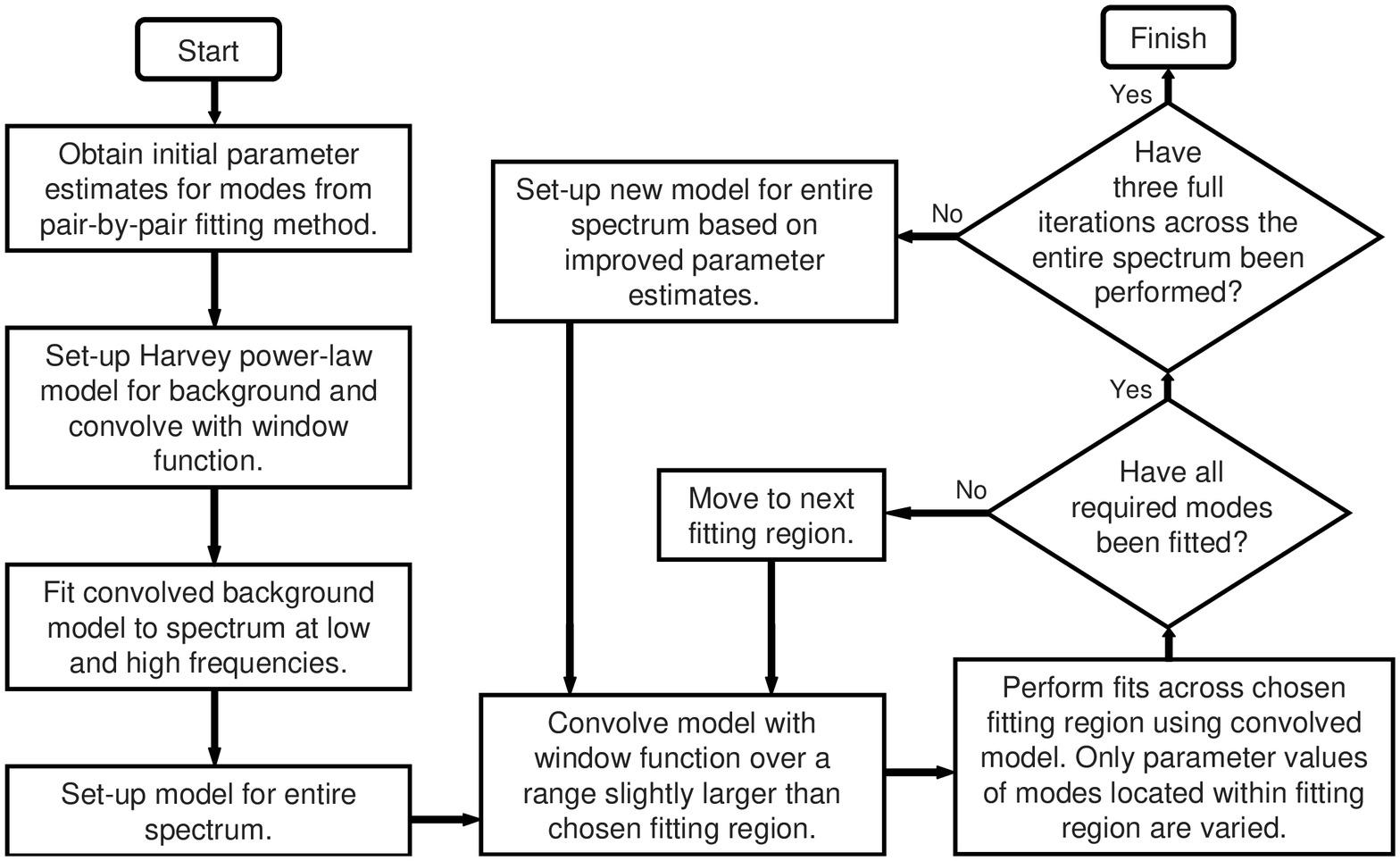}}\caption{Step-by-step
flow diagram representing the pseudo-global fitting algorithm when
fitting gap-affected data.} \label{PG_FlowChart}
\end{figure}

\section{Data}

We have performed a comparative analysis of the PPM and PGM
described above using both real and simulated data. The real data
consists of ground-based observations taken by the Birmingham Solar
Oscillations Network (BiSON). The simulated data were created using
the second-generation solarFLAG simulator described in
\cite{Chaplin2008}.

When using simulated data a common approach is to perform a full
Monte-Carlo analysis whereby many independent realizations of the
same underlying data are fitted. This enables one to take the
average of the fitted mode parameters and compare with the input
values in order to obtain an estimate of any bias. However, in this
analysis we chose instead to perform fits to a single limit spectrum
(i.e., the spectrum one would obtain in the limit of summing an
infinite number of independently generated spectra).
\cite{Toutain2005} showed that fitting the limit spectrum and
comparing the results with the known input values gives a direct
measure of any bias without the need to fit many independent spectra
(see Paper II for more justification of this approach).

The simulated data were created in the same manner as explained in
Paper II using the second-generation solarFLAG simulator described
in \cite{Chaplin2008}. As the main focus of this paper is to test
the benefits of the PGM when applied to data with significant gaps,
a BiSON-like window function was imposed on the simulated data.
Since the simulated limit spectrum is actually created in the
frequency domain, this had to be done by convolving with the
spectral window function (as opposed to simply multiplying the time
series by the window function).

When fitting the real data a 3456-day BiSON time series was used.
The data were collected between December 1998 and July 2008, and the
fill is about 80 percent.

\section{Results}\label{SecResults}

In this section we first give the results of fits to the simulated
data before giving the results for the real data. In both cases we
compare the fitted parameters returned from both the PPM and PGM.

\begin{figure}
\centerline{\includegraphics[width=5.2in]{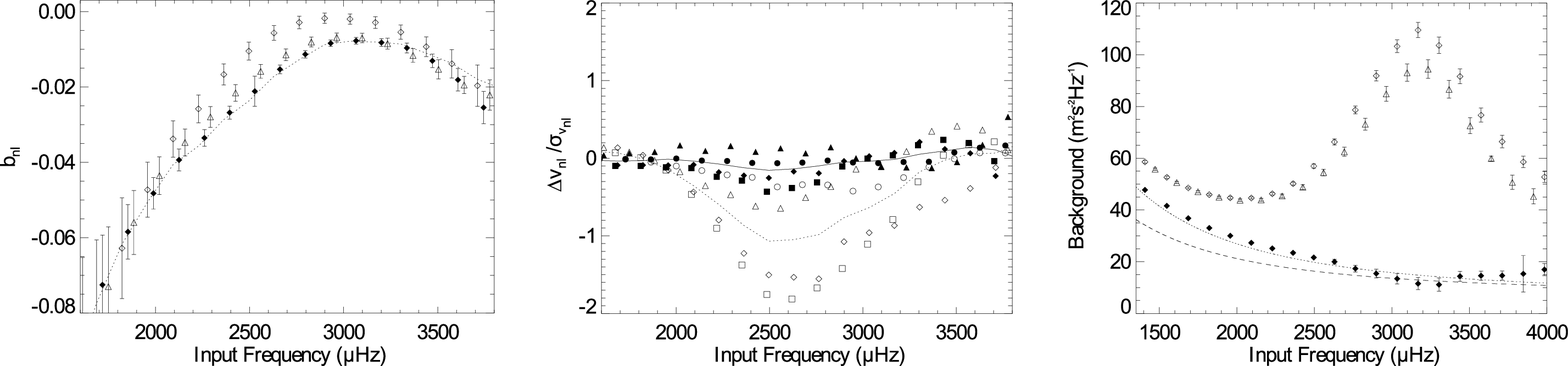}}
\caption{Results of fitting simulated limit spectra convolved with a
BiSON-like spectral window. The left panel gives the fitted
asymmetries with the dotted line giving the input asymmetries. The
middle panel gives the differences between the fitted and input
frequencies as described in the text while the solid and dashed
lines give the means over four points (one of each degree). The
right panel gives the fitted backgrounds with the dashed line giving
the input background and the dotted line the predicted background
(see text). In all cases the open symbols give the results from the
PPM and the black symbols give the results from the PGM. Diamonds
signify $\ell$ = 0 modes, triangles $\ell$ = 1, squares $\ell$ = 2
and circles $\ell$ = 3.} \label{SimData}
\end{figure}

Fig.~\ref{SimData} shows the estimated asymmetries, frequencies and
background when fitting simulated data with an imposed BiSON-like
window function. The asymmetries returned by the PPM clearly give
significantly biased estimates compared with the known input values
as given by the solid line. This result can be explained by the fact
that asymmetric modes lying outside a particular fitting region will
introduce a different amount of excess power at the low frequency
end of the fitting region compared with the high frequency end.
Since this is not accounted for by the model in the PPM, the fitted
asymmetry will be biased. In contrast the fits from the PGM, given
by the black symbols, are considerably more accurate since the
excess power \emph{is} accounted for.

For the PPM the fact that the asymmetries are biased also leads to
bias in the fitted frequencies. This is shown in the middle panel of
Fig.~\ref{SimData}, where, in order to give a direct measurement of
the significance of the bias we have plotted the difference between
the fitted values and the known input values (in the sense fitted -
input) and divided by the estimated uncertainties,
$\sigma_{\nu_{nl}}$.

The right-hand panel of Fig.~\ref{SimData} shows the fitted
backgrounds and it is clear that the PPM return values that
overestimate the true background, as given by the dashed line, by a
considerable amount. In contrast the background fits from the PGM
match much more closely, although some overestimation still remains.
However, there is a very good agreement with the predicted
background one obtains by fitting the power spectrum at low and high
frequencies. The reason why there is a difference between the input
and predicted background is explained in Paper II.

As would be expected, all these results are similar to what was
obtained when fitting the full-fill data given in Paper II. The main
difference is that the biases seen when applying the PPM are more
significant. In the case of the background this is fairly
straightforward to explain. The model in the PPM does not account
directly for the power from modes lying outside the fitting regions.
Therefore, the background is increased to account for the excess
power. In the case of gap-affected spectra there is also excess
power that is taken out of the mode peaks. Within the fitting
regions this is accounted for by convolving the model with the
spectral window. Outside the fitting regions, however, this extra
power is unaccounted for and this will have a small but cumulative
impact inside the fitting window. Therefore, the fitted background
is increased even further to compensate. The PGM removes this
problem by convolving the full spectrum model with the spectral
window.

\begin{figure}
\centerline{\includegraphics[width=5.2in]{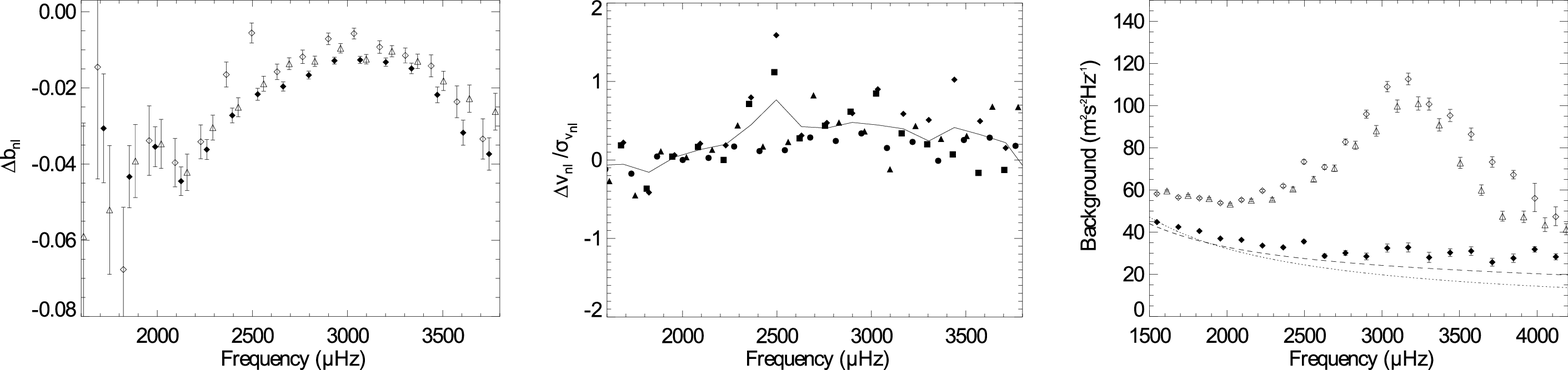}}
\caption{Results of fitting real 3456-day BiSON data. The left panel
gives the fitted asymmetries. The middle panel gives the differences
between the fitted frequencies returned by the two routines (in the
sense PGM minus PPM) divided by the combined formal uncertainties
with the solid line giving the mean over four points (one of each
different degree). The right panel gives the fitted backgrounds with
the dotted line giving the predicted background using physical
fitting values and the dotted line giving the predicted background
using non-physical fitting values. The symbols have the same
meanings as in Fig~\ref{SimData}} \label{BiSONData}
\end{figure}

In Fig.~\ref{BiSONData} we show the fitted asymmetries, frequencies
and backgrounds for the 3456-day BiSON data set. For the asymmetry
and background results we have plotted the same quantities as in
Fig.~\ref{SimData}, whereas for the frequencies we have plotted the
differences between the results returned from the two techniques (in
the sense PGM minus PPM).

The plots show similar trends to those seen for the simulated data,
with the estimates from the PGM being significantly different
compared with those from the PPM. However, in the case of the
frequencies the differences do not appear to be as significant. This
is most likely due to the fact that the modes in the real data do
not perfectly match the fitting model.

The background estimates from the PGM, while still far better than
those from the PPM, do not match the predicted background (i.e., the
background estimated from a fit to the spectrum at low and high
frequencies) as well as in the simulated data. The dotted line in
the background plot represents the predicted background when forcing
the fitting model to take physical (i.e, non negative) values. The
dashed line is the predicted background when this restriction is not
enforced. The fact there is a large difference between the two would
seem to suggest that there may be some non-solar background source
that is dependent on frequency that is not currently being included
in the fitting model, (hence why a less physical model can currently
more accurately represent the background). However, whichever
background prediction is used the fits still lie above it, which
again suggests we do not have a perfect model for the modes.

\section{Summary}

We have modified our pseudo-global fitting method (PGM) outlined in
Paper II to enable the fitting of spectra created from time series
with significant gaps. By using simulated data it was shown that the
PGM still enabled less biased estimates of the mode frequencies and
asymmetries and the underlying background despite the presence of
gaps in the data.

Fits to real BiSON data showed similar difference in the returned
parameter values between the PGM results and pair-by-pair fitting
method (PPM) results as was seen in the simulated data. Given what
was seen when fitting the simulated data we would expect the results
returned from the PGM to be more accurate. Hence we intend to use
the new PGM to fit a long stretch of BiSON data and publish a new
set of updated BiSON frequencies.

Differences between the fitted frequencies from the PPM and PGM were
not as significant as was seen for the simulated data. Also, in the
real data the background fits from the PGM did not match the
predicted background (i.e., the background estimated from a fit to
the spectrum at low and high frequencies) as well as in the
simulated data. These observation suggests that the fitting model
for the modes is still not perfect.

\acknowledgements 
STF acknowledges the support of the Faculty of Arts, Computing,
Engineering and Science (ACES) at the University of Sheffield and
the support of the Science and Technology Facilities Council (STFC).
We also thank all those associated with BiSON which is funded by the
STFC.


\end{document}